\documentclass[12pt,twocolumn,preprint]{aastex6}
\usepackage{CJK}
\usepackage{color}

\usepackage{natbib}
\usepackage{hyperref}
\usepackage{cleveref}
\usepackage{url}

\def\esym{$E_{\rm{sym}}(\rho)$~}
\begin{document}
\title{Constraints on the muon fraction and density profile in neutron stars}
\author{Nai-Bo Zhang\altaffilmark{1} and Bao-An Li\altaffilmark{2}$^{*}$}
\altaffiltext{1}{Shandong Key Laboratory of Optical Astronomy and Solar-Terrestrial Environment, School of Space Science and Physics, Institute of Space Sciences, Shandong University,  Weihai, Shandong, 264209, China\\}
\altaffiltext{2}{Department of Physics and Astronomy, Texas A$\&$M University-Commerce, Commerce, TX 75429, USA\\
\noindent{$^{*}$Corresponding author: Bao-An.Li@Tamuc.edu}}

\begin{abstract}
Muons in neutron stars (NSs) play especially important roles in addressing several interesting new physics questions associated with detecting as well as understanding interactions and astrophysical effects of muonphilic dark matter particles. The key model inputs for studying the latter are the total muon mass $M_{\mu}$, the muon mass fraction $M_{\mu}/M_{\rm NS}$ over the NS mass $M_{\rm NS}$ and the muon radial density profile $\rho_{\mu}(r)$ in NSs of varying masses. We investigate these quantities within a minimum model for the core of NSs consisting of neutrons, protons, electrons, and muons using an explicitly isospin-dependent parametric Equation of State (EOS) constrained by available nuclear laboratory experiments and the latest astrophysical observations of NS masses, radii and tidal deformabilities. We found that the absolutely maximum muon mass $M_{\mu}$ and its mass fraction $M_{\mu}/M_{\rm NS}$ in the most massive NSs allowed by causality are about 0.025 $M_\odot$ and 1.1\%, respectively. For the most massive NS of mass 2.14 $M_\odot$ observed so far,  they reduce to about 0.020 $M_\odot$ and 0.9\%, respectively. We also study respective effects of individual parameters describing the EOS of high-density neutron-rich nucleonic matter on the muon contents in NSs with varying masses. We found that the most important but uncertain nuclear physics ingredient for determining the muon contents in NSs is the high-density nuclear symmetry energy.
\end{abstract}
\maketitle

\section{Introduction}
Neutron stars (NSs) have long been recognized as useful cosmic laboratories to capture and study properties of Dark Matter (DM) particles while their direct and indirect detections in terrestrial laboratories remain elusive, see, e.g., Refs. \citep{DM-review,Das20} for recent reviews. In particular, muons in NSs were found to play a special role in addressing several interesting new questions and anomalies in particle physics beyond the Standard Model as well as the search for DM and new mesons mediating possible new forces in nature at various length and/or energy scales. For example, assuming NSs are made of neutrons, protons, electrons, and muons (i.e, the $npe\mu$ model for NSs) and a single collision between the DM particle and a particle in NSs is required for the DM particle to become gravitationally bound in NSs, the DM-muon scattering was found to have a greater sensitivity than all other techniques across almost the entire mass range for all types of interactions considered \citep{DM-review}, leading to some special research interests in muonphilic DM particles that interact only with muons. Indeed, within some new physics models for muonphilic DM motivated by existing anomalies in the muon sector, such as the anomalous magnetic moment and the LHCb's indication for the lepton-flavor nonuniversality in $B\rightarrow K\mu^+\mu^-$ decays \citep{Garani19}, the unique nongravitational interaction between muonphilic DM particle and muons as well as DM annihilations in NSs leads to an appreciable heating of some old and cold NSs, which can be potentially observed by the infrared telescopes \citep{DM-review,Garani19}. Interestingly, the sensitivity was predicted to be orders of magnitude more powerful than the current DM-electron scattering bounds from terrestrial experiments \citep{DM-review}. Moreover, long-range muonic forces mediated by proposed new mesons may result in a dipole radiation as well as a modification of the chirp mass during the inspiral phase of merging NS binaries if they both contain significant amounts of muons \citep{Dror19}.

All of the predicted muonphilic DM capture cross sections and associated effects on properties of both isolated NSs and their mergers depend sensitively on the muon mass fraction $M_{\mu}/M_{\rm NS}$ and its density profile $\rho_{\mu}(r)$ in NSs. It is well known that muons start to appear in NSs when the electron chemical potential becomes higher than the muon rest mass of $m_{\mu}c^2=105.66$ MeV. Moreover, it is also well known that the electron chemical potential in NSs at $\beta$ equilibrium is uniquely determined by the nuclear symmetry energy $E_{\rm{sym}}(\rho)$ at baryon density $\rho$. In particular, the high-density behavior of $E_{\rm{sym}}(\rho)$ determines the critical density above which muons start to appear \citep[see, e.g., ][]{Tesym} for a collection of reviews. Therefore, both the mass fraction and density profiles of muons in NSs are expected to be strongly affected by the high-density behavior of nuclear symmetry energy. While the nuclear symmetry energy especially at supra-saturation densities remains the most uncertain part of the Equation of State (EOS) of dense neutron-rich nucleonic matter, significant constraints on the $E_{\rm{sym}}(\rho)$ have been obtained recently from both terrestrial experiments and astrophysical observations. In particular, analyses of NS radii from earlier Chandra \citep{Steiner10} X-ray data as well as the tidal deformability of NSs from GW170817 \citep{LIGO18} together with the latest NS maximum mass $M=2.14^{+0.10}_{-0.09}$~$M_\odot$ from observations of PSR~J0740+6620 \citep{Mmax} have facilitated the recent extraction of tighter constraints on the EOS of dense neutron-rich nucleonic matter, especially its symmetry energy term \citep[see, e.g., ][]{Zhang18,Zhang19,Zhang19a,Zhang19b,Nakazato19,Baillot19,LWChen19,YZhou19}.

The main purpose of this work is to examine the maximum muon mass fractions and density profiles in NSs in the high-density EOS parameter space allowed by the available nuclear laboratory experiments and the latest observations of NS properties (mass, radius, and tidal deformability) including the simultaneous measurement of both the mass and radius of PSR J0030+0451 by the Neutron Star Interior Composition Explorer (NICER) Collaboration \citep{Riley2019,Raa,Bil,Miller2019,Bog1,Bog2,Gui}. The results of this work can be used as constrained model inputs in future studies of the new physics associated with the interactions between muonphilic DM and muons in NSs.

The manuscript is arranged as follows: In the next section, we first outline the $npe\mu$ model used in our previous work \citep{Zhang18,Zhang19,Zhang19a,Zhang19b,Xie19}
and present work for inverting NS observables using a parametric EOS with explicit isospin dependence. Then effects of high-density EOS parameters on the muon mass fractions will be examined in Section \ref{effects}. In Section \ref{EOS-space}, we update the allowed high-density EOS parameter space by the latest NS observations.
In Section \ref{Muon-constraints}, we report the muon mass fractions and density profiles in canonical NSs, NSs with a mass of 2.14 $M_{\odot}$ as well as NSs on the causality surface where the speed of sound is the same as the speed of light and discuss their dependences on the remaining uncertainties of high-density EOS of neutron-rich matter.
Finally, a summary and outlook will be given.

\section{The minimum model for calculating the muon contents in neutron stars}\label{model}
In the following, we first recall the main features and equations of the $npe\mu$ model of NSs widely used in the literature to ease the discussions of our results. We emphasize that this is the minimum model for studying muons in NSs. It is well known that hyperons and baryon resonances are predicted to appear above certain critical densities   \citep[see, e.g., ][]{Isaac1,Isaac2,Pro19,Ramos}. In particular, negatively charged particles, such as $\Sigma^-$ and $\Delta^-(1232)$ are among the first to be formed as the density increases from the crust to the core of NSs. Their appearances will affect significantly the fraction of other negatively charged particles, such as electrons and muons. However, there are currently large uncertainties associated with the production mechanisms and interactions of both hyperons and baryon resonances in NSs. For instance, the $\Delta^-(1232)$ may appear above a critical density as low as $\rho_0$ depending on its completely unknown coupling strength with the $\rho$-meson \citep[see, e.g.,][]{D1,Cai,AngLi,Kim,Sahoo,Armen1,Ramos}.
Thus, in some models the muon fraction could decrease very quickly to zero at high densities once the formation of $\Sigma^-$ and/or $\Delta^-(1232)$ is considered. In addition, the phase transition from nuclear matter to quark matter will also suppress the formation of muons \citep{Weber05}, and muons may not appear in quark stars \citep{Masuda13}. Thus, a more accurate study of muon contents requires a much more comprehensive understanding about the composition and phases of dense matter beyond the abilities of the $npe\mu$ model of NSs. Nevertheless, given the largely exploratory nature of studies about the muonphilic DM in NSs in the literature, the most optimistic estimate of the muon contents in NSs within the minimum model is presently sufficient.

Our EOS model for NSs has three parts. For nucleons in the core, as we shall discuss in detail next we use an explicitly isospin-dependent parametric EOS.
The core EOS is then connected smoothly at the core-crust transition point to the NV EOS \citep{Negele73} for the inner crust
followed by the BPS EOS  \citep{Baym71} for the outer crust. Moreover, as discussed in detail in our previous work in Ref. \citep{Zhang18}, when the core EOS parameters are varied, the core-crust transition density and pressure are calculated consistently.

Within the $npe\mu$ model for the core of NSs, the core pressure $P(\rho, \delta)=\rho^2\frac{d\epsilon(\rho,\delta)/\rho}{d\rho}$
can be obtained from the total energy density $\epsilon(\rho, \delta)$ of NS matter with nucleon isospin asymmetry $\delta=(\rho_{\rm{n}}-\rho_{\rm{p}})/\rho$
at baryon density $\rho$
\begin{equation}\label{energydensity}
  \epsilon(\rho, \delta)=\epsilon_b(\rho, \delta)+\epsilon_l(\rho, \delta),
\end{equation}
where $\epsilon_b(\rho, \delta)$ and $\epsilon_l(\rho, \delta)$ are the energy density of baryons and leptons, respectively. The $\epsilon_b(\rho, \delta)$ can be calculated from
\begin{eqnarray}\label{Ebpa}
  \epsilon_b(\rho,\delta)=\rho E(\rho,\delta)+\rho M_N,
\end{eqnarray}
where $E(\rho,\delta)$ is the energy per nucleon (nucleon specific energy) of asymmetric nuclear matter and $M_N$ is the average nucleon mass. The $\epsilon_l(\rho, \delta)$ is obtained from
\begin{equation}\label{el}
  \epsilon_l(\rho, \delta)=\eta\phi(t)
\end{equation}
with
\begin{equation}
  \eta=\frac{m_l^4}{8\pi^2},~~
  \phi(t)=t\sqrt{1+t^2}(1+2t^2)-ln(t+\sqrt{1+t^2}),
\end{equation}
and
\begin{equation}
  t=\frac{(3\pi^2\rho_l)^{1/3}}{m_l}
\end{equation}
is taken from the noninteracting Fermi gas model ($\hbar=c=1$) \citep{Oppenheimer39}. The chemical potential of particle $i$ is given by
\begin{equation}\label{chemicalnpe}
  \mu_i=\frac{\partial\epsilon(\rho,\delta)}{\partial\rho_i}.
\end{equation}
The relative particle fractions at different densities in NSs are obtained through the chemical ($\beta$) equilibrium condition
$
  \mu_n-\mu_p=\mu_e=\mu_\mu
$
and the charge neutrality condition
$  \rho_p=\rho_e+\rho_\mu
$
for the density of proton, electron, and muon, respectively.
The condition $\mu_e=\mu_\mu$ can be written explicitly as
\begin{equation}
m_e \left[1+\frac{(3\pi^2 \rho_e)^{2/3}}{m_e^2}\right]^{1/2} = m_\mu \left[1+\frac{(3\pi^2 \rho_\mu)^{2/3}}{m_\mu^2}\right]^{1/2}
\end{equation}
where $m_e$ is the electron mass, thus relating the number densities of muons and electrons through \citep{Dror19}
\begin{equation}
\rho_\mu = \frac{m_e^3}{3\pi^2}\left[1+\frac{(3\pi^2 \rho_e)^{2/3}}{m_e^2}-\frac{m_\mu^2}{m_e^2}\right]^{3/2}.
\end{equation}
Then, the total muon rest mass in the NS local frame can be calculated from the product of $m_\mu$ and the total number of muons, i.e., 
\begin{equation}
M_\mu = m_\mu\cdot 4\pi \int_0^R \rho_\mu r^2 dr.
\end{equation}
While one can also integrate the muon energy density given in Eq. (\ref{el}) to obtain the total muon gravitational mass as one does for calculating the NS gravitational mass M$_{\rm NS}$, we present in this paper 
the total muon rest mass in the NS local frame as it gives directly the total number of muons that muonphilic DM particles can interact with inside the NS considered. The total number of muons is the most basic and useful information for studying interactions of muonphilic DM particles and their effects on properties of NSs. Moreover, the total muon rest mass and gravitational mass have very similar dependences on the high-density EOS parameters. While our quantitative conclusions in this work are the same by studying either the total muon gravitational mass or rest mass, there are maybe interesting questions where it is important to know the momentum distribution of muons and the corresponding kinetic contribution of muons to the total gravitational mass of NSs. Then, the total muon gravitational mass can be calculated within the same framework presented here.

To obtain fractions of neutrons, protons, electrons, and muons relative to the total baryon density $\rho$ as well as the pressure in NSs at $\beta$ equilibrium, one has to calculate the difference in chemical potentials of neutrons and protons $\mu_n-\mu_p$ from the nucleon specific energy \citep{Bom91}
\begin{equation}\label{PAEb}
  E(\rho,\delta)\approx E_0(\rho)+E_{\rm{sym}}(\rho)\delta^2
\end{equation}
where $E_0(\rho)$ is the nucleon specific energy in symmetric nuclear matter (SNM) and $E_{\rm{sym}}(\rho)$ is the nuclear symmetry energy. As discussed in detail in our previous work \citep{Zhang18,Zhang19,Zhang19a,Zhang19b,Xie19}, we parameterize both the $E_0(\rho)$ and $E_{\rm{sym}}(\rho)$ up to the third power of $[(\rho-\rho_0)/3\rho_0]$ as
\begin{eqnarray}\label{E0para}
  E_{0}(\rho)&=&E_0(\rho_0)+\frac{K_0}{2}(\frac{\rho-\rho_0}{3\rho_0})^2+\frac{J_0}{6}(\frac{\rho-\rho_0}{3\rho_0})^3,\\
  E_{\rm{sym}}(\rho)&=&E_{\rm{sym}}(\rho_0)+L(\frac{\rho-\rho_0}{3\rho_0})+\frac{K_{\rm{sym}}}{2}(\frac{\rho-\rho_0}{3\rho_0})^2\nonumber\\
  &+&\frac{J_{\rm{sym}}}{6}(\frac{\rho-\rho_0}{3\rho_0})^3\label{Esympara}.
\end{eqnarray}

\begin{figure*}
\begin{center}
\resizebox{0.8\textwidth}{!}{
  \includegraphics{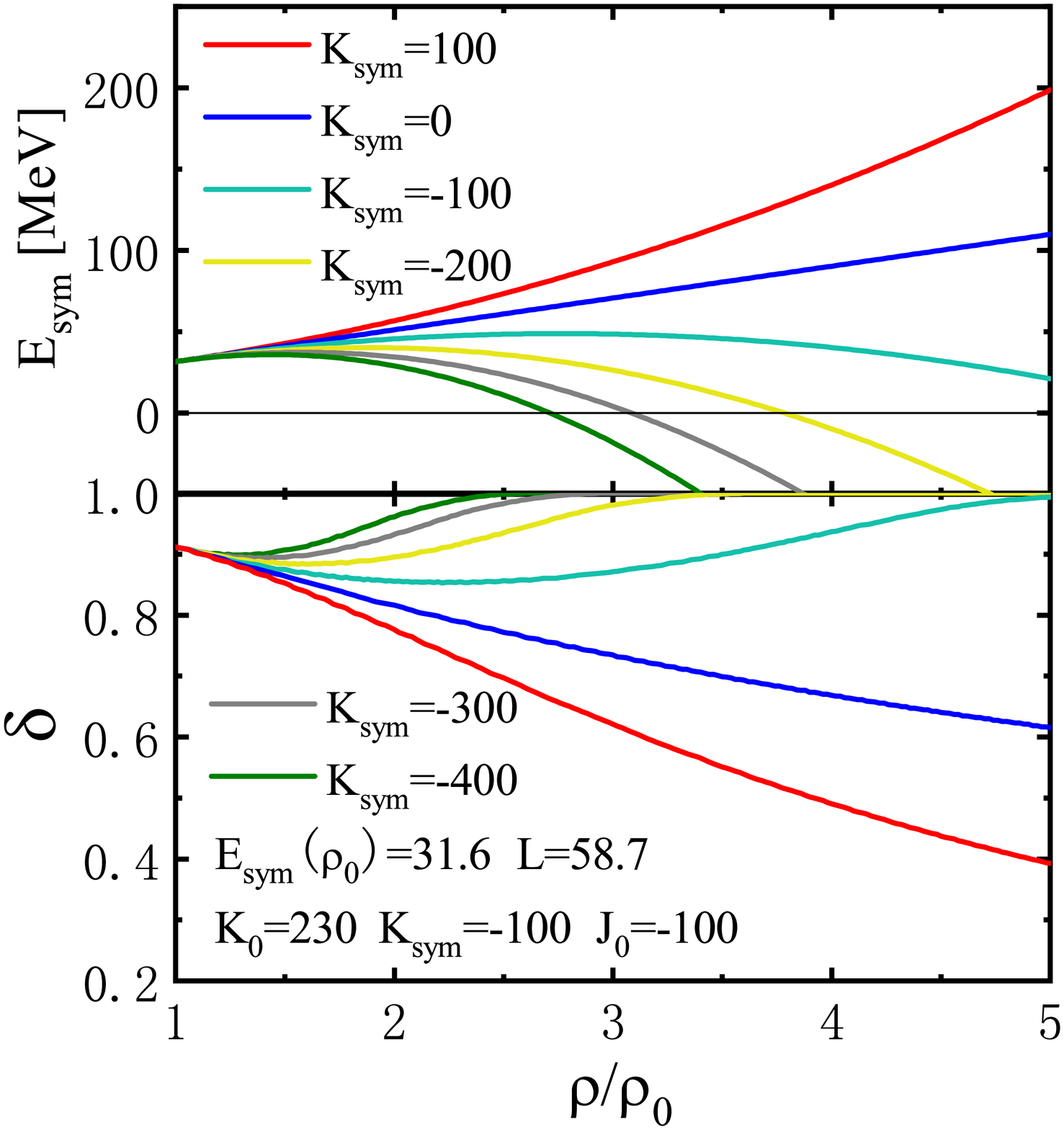}
  \includegraphics{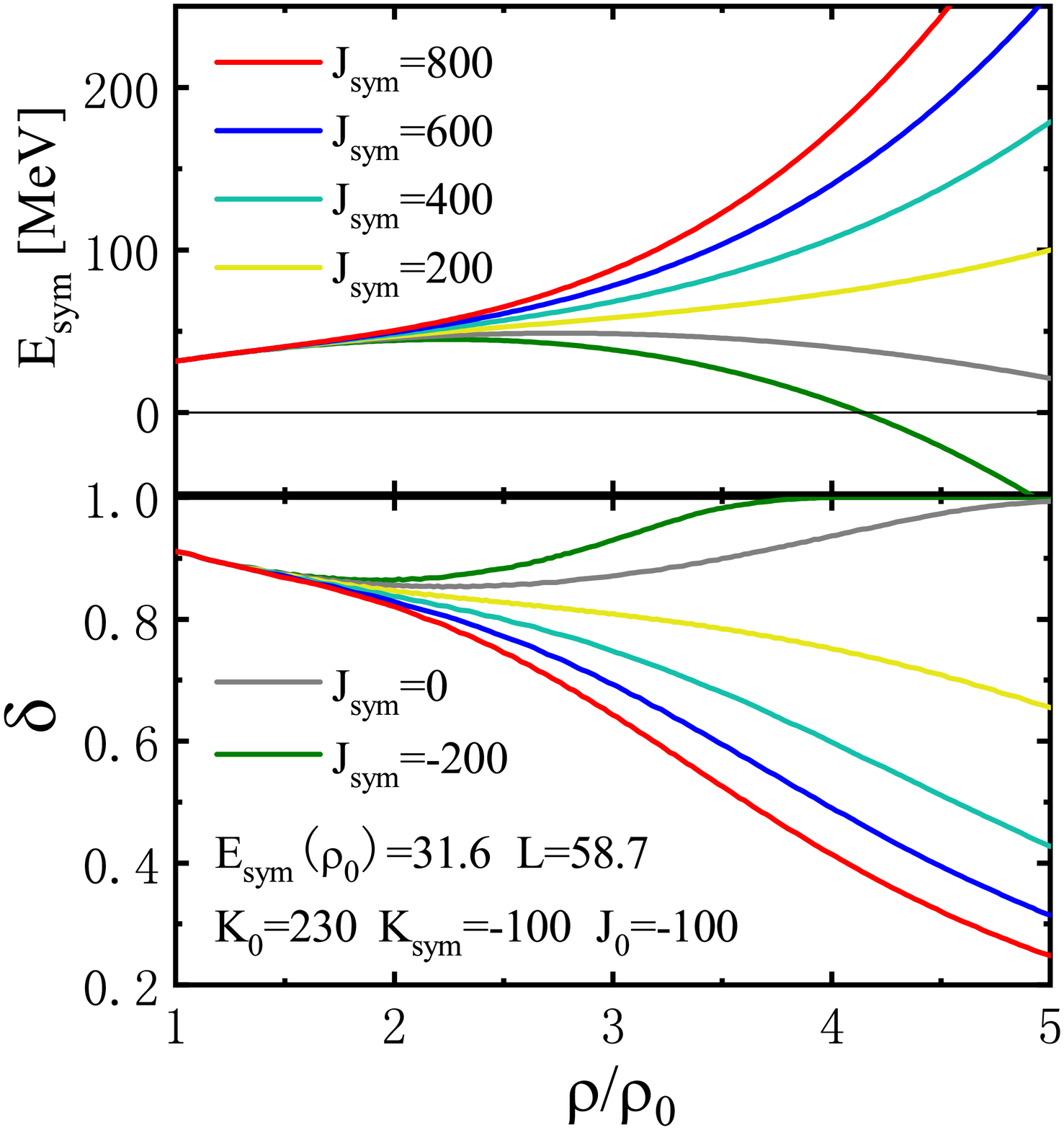}
  }
  \caption{The symmetry energy $E_{\rm{sym}}(\rho)$ and isospin asymmetry profile $\delta(\rho)$ in NS matter at $\beta$ equilibrium as a function of the reduced density $\rho/\rho_0$ for $K_{\rm{sym}}=-400$, -300, -200, -100, 0, and 100 MeV (left), and $J_{\rm{sym}}=-200$, 0, 200, 400, 600, and 800 MeV (right), respectively, while all other parameters are fixed as the specified values. Taken from \citet{Zhang18}}\label{Ksymeffect}
  \end{center}
\end{figure*}

Considering the above parameterizations as empirical energy density functionals, the parameters $K_0$ and $J_0$ are the incompressibility and skewness of SNM at saturation density $\rho_0$, while the $L$, $K_{\rm{sym}}$ and $J_{\rm{sym}}$ are the slope, curvature and skewness of nuclear symmetry energy at $\rho_0$.
Previous analyses of many kinds of terrestrial nuclear experiments and astrophysical observations have constrained the values of  $E_0(\rho_0)$, $K_0$, $E_{\rm sym}(\rho_0)$ and $L$ to around
 $E_0(\rho_0)=-15.9 \pm 0.4$ MeV \citep{Brown14}, $K_0\approx 240 \pm 20$ MeV \citep{Shlomo06,Piekarewicz10}, $E_{\rm sym}(\rho_0)=31.7\pm 3.2$ MeV and $L\approx 58.7\pm 28.1 $ MeV \citep{Li13,Oertel17}, respectively. However, the three parameters $K_{\rm{sym}}$, $J_{\rm{sym}}$, and $J_0$ characterizing the high-density EOS of neutron-rich matter are only known roughly in the range of $-400 \leq K_{\rm{sym}} \leq 100$ MeV, $-200 \leq J_{\rm{sym}}\leq 800$ MeV, and $-800 \leq J_{0}\leq 400$ MeV \citep{Tews17,Zhang17}, respectively. These parameters together thus span a large high-density EOS parameter space. In particular, the $K_{\rm{sym}}$ and $J_{\rm{sym}}$ together control the high-density behavior of nuclear symmetry energy. The latter determines uniquely the proton fraction (i.e., the total lepton fraction) of NSs at $\beta$ equilibrium through the condition $\mu_n-\mu_p=\mu_e=\mu_\mu\approx 4\delta E_{\rm{sym}}(\rho)$ when the chemical potential of electrons is higher than the rest mass of muons above certain densities. One thus expects the $K_{\rm{sym}}$ and $J_{\rm{sym}}$ parameters to play significant roles in determining the muon contents. Of course, effects of the three high-density EOS parameters on the mass, radius and muon contents are correlated and have to be examined together systematically. It is thus useful to recall here typical effects of the $K_{\rm{sym}}$ and $J_{\rm{sym}}$ parameters on the density profile of isospin asymmetry $\delta(\rho)$ in NSs at $\beta$ equilibrium in Figure \ \ref{Ksymeffect}.  In the left panel, we show results with $K_{\rm{sym}}=-400$, -300, -200, -100, 0, and 100 MeV, while in the right panel $J_{\rm{sym}}=-200$, 0, 200, 400, 600, and 800 MeV, with all other parameters fixed at the values specified, respectively. Clearly, as the \esym varies broadly at high-densities by varying either the $K_{\rm{sym}}$ or $J_{\rm{sym}}$ parameter, the resulting density profile of isospin asymmetry $\delta(\rho)$ changes accordingly between 0 (symmetric nuclear matter) and 1 (pure neutron matter).

Within the $K_{\rm sym}-J_{\rm sym}-J_0$ 3D high-density EOS parameter space with $E_0(\rho_0)$, $K_0$, $E_{\rm sym}(\rho_0)$ and $L$ fixed at their currently known most probable empirical values described above, we can solve efficiently some NS inverse-structure problems. Namely, given a value of a NS observable, e.g., radius, mass, tidal deformability, moment of inertia, etc, we can find all necessary combinations of the high-density EOS parameters to reproduce the specified observable, see, e.g., Ref. \citep{Zhang19} for details. We shall first explore the individual effects of the three high-density EOS parameters within the allowed EOS space discussed above on the muon fraction in NSs of varying masses. We then study how the latest astrophysical observations further limit the high-density EOS parameter space and the resulting consequences on the muon contents in NSs. 

\begin{figure*}
  \centering
   \resizebox{0.9\textwidth}{!}{
  \includegraphics{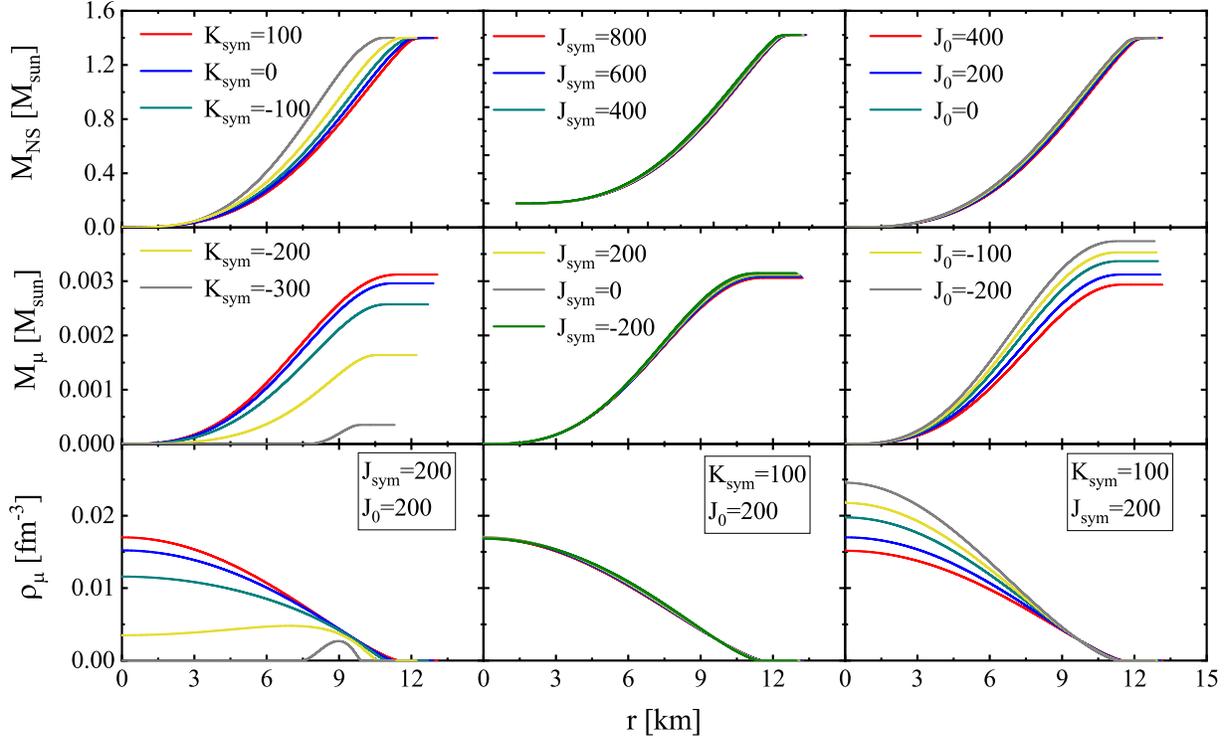}
  }
  \caption{The profile of NS mass (up panels), muon mass (middle panels), and muon density (lower panels) for canonical NSs with a mass of 1.4 $M_\odot$. The effects of high-density EOS parameters $K_{\rm sym}$ (left panels), $J_{\rm sym}$ (middle panels), and $J_0$ (right panels) are also shown.}\label{M14}
\end{figure*}

\section{Effects of individual high-density EOS parameters on muon contents in neutron stars}\label{effects}
It is known from previous studies that the SNM EOS dominates the masses while the symmetry energy dominates the radii of NSs, see, e.g., Ref. \citep{LiSteiner}. Thus, the three high-density EOS parameters are expected to play different roles in determining the muon contents in NSs. Here we first fix the NS mass at 1.4 $M_\odot$ without any restriction on its radius. Shown in Figure\ \ref{M14} are the accumulated NS mass (upper panels) and muon mass (middle panels) up to the radial coordinate $r$ from the center as well as the muon radial density profile (lower panels) up to the radius $R$ where the pressure becomes zero for canonical NSs with a total mass of 1.4 $M_\odot$. As shown in the upper raw, the integrated mass saturates at 1.4 $M_\odot$ by design but the radius $R_{1.4}$ is between about 11.5 km and 13 km when the three parameters are varied. In particular, the radius $R_{1.4}$ increases significantly with the increasing value of $K_{\rm sym}$ but increases only slightly with the increasing values of $J_{\rm sym}$ and $J_0$. This finding is consistent with our prior knowledge in the literature, see, e.g., Ref. \citep{BALI19}, for a comprehensive review. The overall average density in a 1.4 $M_\odot$ NS is in the range of $(1.23\sim 1.97)\rho_0$ \citep{WK}. The nuclear symmetry energy in this density range is mainly controlled by the $K_{\rm sym}$ while the $J_{\rm sym}$ becomes more influential at higher densities as shown in Figure\ \ref{Ksymeffect}. Focusing on effects of the $K_{\rm sym}$ by fixing the $J_{\rm sym}$ and $J_0$ parameters in the left panel, it is seen that higher values of $K_{\rm sym}$ lead to higher masses of muons. Moreover, most of these muons reside in the interior of NSs. The observed effects of $K_{\rm sym}$ on the muon mass and profile are easily understandable by inspecting the corresponding symmetry energy and isospin asymmetry profile $\delta(\rho)$ at $\beta$ equilibrium shown in Figure\ \ref{Ksymeffect}. Higher values of $K_{\rm sym}$ make the symmetry energy stiffer, thus energetically favor the formation of a more neutron-poor system. It is mainly due to the $E_{\rm sym}\delta^2$ term in the EOS of Eq. (\ref{PAEb}) and is known generally as the isospin fractionation in the literature, see, e.g., Refs. \citep{Muller95,Xu00,ditoro,LCK08}. The resulting higher proton fraction makes the electron chemical potential higher, facilitating more muon production in NSs. It is also interesting to note that with very low values of $K_{\rm sym}$, e.g.,  $K_{\rm sym}=-300$ MeV, as shown in Figure\ \ref{Ksymeffect}, the symmetry energy becomes super-soft at high densities. Consequently, it is energetically favorable for NS matter to be pure neutron matter at high densities. In this case, only for densities less than about $3\rho_0$, the symmetry energy is stiff enough for
protons (thus leptons) to appear. Therefore, as shown in the lower window of the left panel of Figure \ref{M14}, muons only appear at low densities towards the surface of NSs for $K_{\rm sym}=-300$ MeV.

On the other hand, fixing both the  $K_{\rm sym}$ and $J_0$, effects of the $J_{\rm sym}$ on the 1.4 $M_\odot$ NS are rather small as shown in the middle column of Figure \ref{M14} while there is a clear tendency that the radius $R_{1.4}$ increases from 12.96 km for $J_{\rm sym}=-200$ MeV to 13.23 km for $J_{\rm sym}=800$ MeV. This is easily understandable as the $J_{\rm sym}$ controls mainly the behavior of nuclear symmetry energy above about $3\rho_0$ while the average density in canonical NSs is significantly less than this density.

\begin{figure*}
  \centering
   \resizebox{0.8\textwidth}{!}{
   \includegraphics{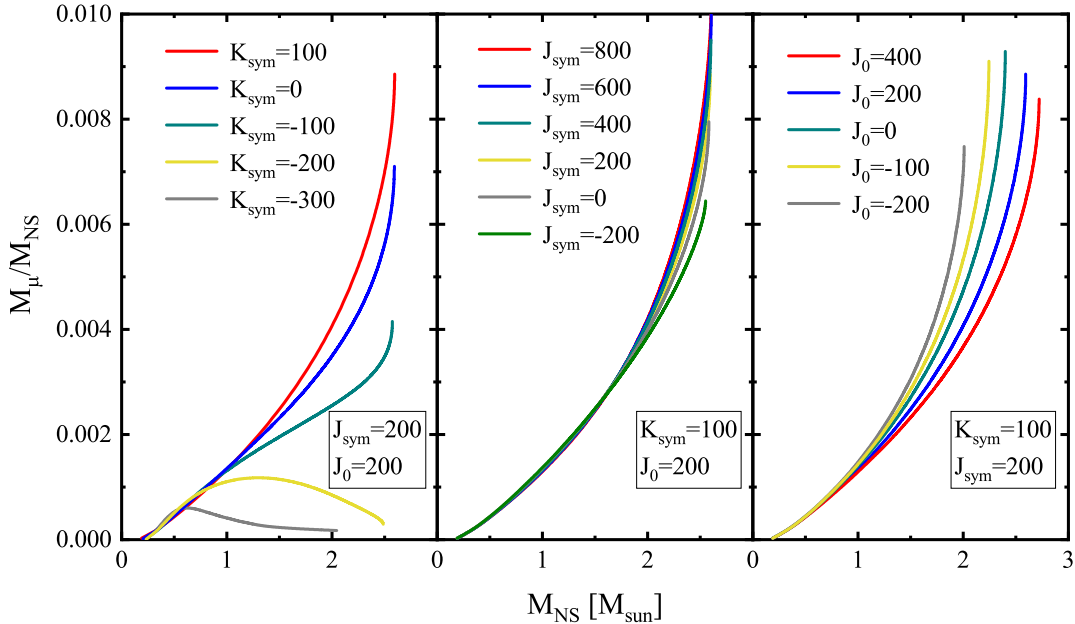}
   }
  \caption{Effects of high-density EOS parameters $K_{\rm sym}$ (left panel), $J_{\rm sym}$ (middle panel), and $J_0$ (right panel) on the muon mass fraction $M_{\mu}/M_{\rm NS}$ as a function of the NS mass $M_{\rm NS}$.}\label{Mmueffects}
\end{figure*}

For comparisons, it is also interesting to examine effects of the stiffness of high-density SNM EOS characterized by the $J_0$ parameter in the right column by fixing the $K_{\rm sym}$ and $J_{\rm sym}$ at values giving a rather stiff symmetry energy at suprasaturation densities. In this case, the functional form of the symmetry energy is fixed. However, the average baryon density and its radial profile are slightly different when varying the $J_0$ parameters. It is seen that the radius $R_{1.4}$ is slighter larger with higher values of $J_0$ as shown in the upper window of the right column of Figure \ref{M14}. Thus, with the NS mass fixed at 1.4 $M_\odot$, the baryon density is appreciably lower throughout the NS with higher values of $J_0$. As the symmetry energy increases with increasing density when both the $K_{\rm sym}$ and $J_{\rm sym}$ are positive, the reduced average baryon density with higher values of $J_0$ makes the proton fraction and the corresponding electron/muon chemical potential lower, leading to a lower production of muons. This effect is clearly shown in the middle and lower windows of the right column. Of course, if we reset the values of both $K_{\rm sym}$ and $J_{\rm sym}$ such that the resulting symmetry energy is very soft, the effects of $J_0$ on muons will be reduced correspondingly. Thus, effects of the three high-density EOS parameters on muons have to be studied by allowing them all to freely change in the whole EOS parameter space allowed by currently available constraints as we shall discuss in Section \ref{Muon-constraints}.

Now we remove the restriction on the NS mass and study respective effects of $K_{\rm sym}$, $J_{\rm sym}$, and $J_0$ on the muon mass fraction $M_{\mu}/M_{\rm NS}$ as a function of the varying NS mass $M_{\rm NS}$ in Figure \ref{Mmueffects}. First of all, it is seen that the three parameters have different effects on the maximum mass of NSs in the range of 2 to 2.7$M_\odot$. As one expects, the $J_0$ has the strongest influence on the NS maximum mass, while extremely soft symmetry energy due to small values of $K_{\rm sym}$ and/or $J_{\rm sym}$
may reduce the maximum mass of NSs supported by the resulting EOS. It is seen from the left panel that the increase of $K_{\rm sym}$ changes the $M_{\mu}/M_{\rm NS}$ ratio significantly for $M>0.8$ $M_\odot$, while effects of $J_{\rm sym}$ are limited for $M<1.6$ $M_\odot$ as the $J_{\rm sym}$ starts to affect the symmetry energy only when the density $\rho$ is higher than about $3\rho_0$ \citep{Zhang19a}.

Generally speaking, a stiffer symmetry energy due to large values of $K_{\rm sym}$ and/or $J_{\rm sym}$ increases both the muon mass and the total NS mass. But their ratio $M_{\mu}/M_{\rm NS}$ may increase or decrease. As the muon mass fraction $M_{\mu}/M_{\rm NS}$ is found to increase with the stiffness of symmetry energy, the resulting increase in the muon mass is apparently larger than that in the total NS mass. It is also very interesting to note that when the $K_{\rm sym}$ values are largely negative (e.g., -200 or -300 MeV), the muon fraction decreases in more massive NSs as shown in the left window of Figure \ref{Mmueffects}. This is again due to the fact that with these negative values of $K_{\rm sym}$ the symmetry energy becomes super-soft favoring the formation of pure neutron matter at densities higher than about $3\rho_0$. Thus, in this case there are less muons in the core of more massive NSs. In principle, largely negative values of $J_{\rm sym}$ have the same effects. This is, however, absent when the $K_{\rm sym}$ has a large positive value as shown in the middle window. Nevertheless, it is seen clearly that when $J_{\rm sym}$ becomes more negative, there are less muons in more massive NSs, which is consistent with the effects of varying the $K_{\rm sym}$ parameter. On the other hand, the increased stiffness $J_0$ of high-density SNM EOS reduces the muon mass as shown in Figure \ref{Mmueffects} but increases significantly the total NS mass. Consequently, the muon mass fraction $M_{\mu}/M_{\rm NS}$ decreases quickly with the increasing value of $J_0$ parameter.

\section{Observational constraints on the high-density EOS parameter space}\label{EOS-space}
In the previous section, we examined the EOS effects on muon contents in NSs within the large uncertainty ranges of the high-density EOS parameters predicted by nuclear theories and allowed by available nuclear laboratory experiments. Constraining the high-density EOS of neutron-rich matter has long been a shared goal of both astrophysics and nuclear physics. Indeed, observations of NSs using several different kinds of messengers in recent years have already provided some useful constraints on the high-density EOS. Within the model framework discussed above, we have recently studied how the three high-density EOS parameters are constrained by NS observations \citep{Zhang18,Zhang19,Zhang19a,Zhang19b,Xie19}. These include the radii $R_{1.4}$ of canonical NSs, e.g., $10.62<R_{\rm{1.4}}< 12.83$ km from analyzing quiescent low-mass X-ray binaries \citep{Lattimer2014},  the dimensionless tidal deformability $70 \leq \Lambda_{1.4}\leq 580$ from the refined analysis of GW170817 data \citep{LIGO18}, and the latest NS maximum mass $M=2.14^{+0.10}_{-0.09}$~$M_\odot$ from the observations of PSR~J0740+6620 \citep{Mmax}. Of course, the causality and dynamical stability conditions have to be satisfied through out the entire NSs. It has been recognized earlier that the available lower limits on both the $R_{1.4}$ and $\Lambda_{1.4}$ are currently outside the crossline between the causality surface and the NS maximum mass of M=2.14 $M_\odot$ in the 3D high-density EOS parameter space. It is also well known that the extraction of the lower limit of $\Lambda_{1.4}$ from the GW170817 data suffers from large uncertainties and the results are significantly model dependent. The lower limits of $R_{1.4}$ and $\Lambda_{1.4}$ extracted so far thus do not provide more tight constraints on the EOS than that set by the causality condition and the maximum mass of NSs, see, e.g., \citet{BALI19} for a recent review. Here we first compile the currently known, most effective and tightest constraints on the $K_{\rm sym}-J_{\rm sym}-J_0$ 3D high-density EOS parameter space. We then examine if and to what degree the simultaneous measurements of NS mass and radius of PSR J0030+0451 reported very recently by the NICER Collaboration \citep{Miller2019,Riley2019} may help further constrain the high-density EOS parameter space. The updated constraints on the high-density EOS space will then be used in the next section to set limits on the muon contents in NSs with different masses.
\begin{figure}
  \centering
  \includegraphics[bb=5 52 555 587, width=9cm]{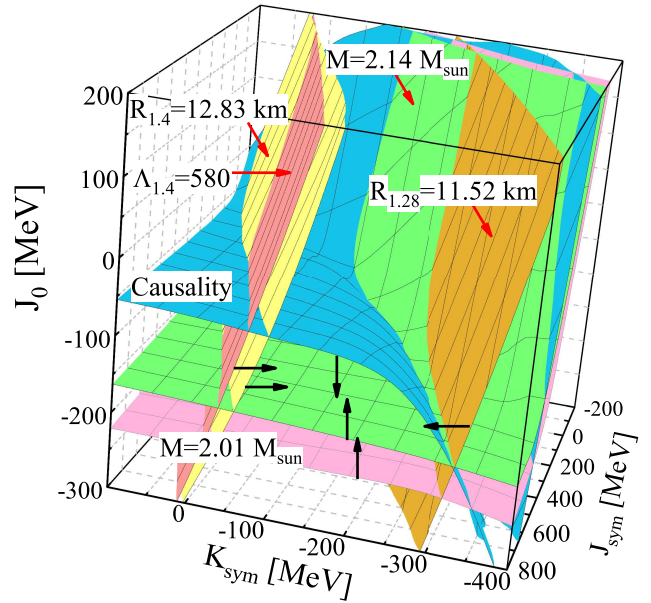}
  \caption{Constant surfaces of neutron star observables with values indicated by the red arrows and the causality condition in the 3D $K_{\rm sym}-J_{\rm sym}-J_0$ EOS parameter space: the NS maximum mass of M=2.14$M_{\odot}$ (green surface) or 2.01$M_\odot$ (pink surface), the radius of canonical NS $R_{1.4}=12.83$ km (yellow surface) or $R_{1.28}=11.52$ km (orange surface) for a NS of mass 1.28$M_{\odot}$, the dimensionless tidal deformability of canonical NS $\Lambda_{1.4}=580$ (red surface), and the causality surface (blue) on which the sound speed equals the speed of light in centers of most massive NSs supported at the point of the EOS parameter space.}\label{M214a}
\end{figure}

Using the direct inversion technique for solving NS inverse-structure problems within the $npe\mu$ model discussed in detail in Refs. \citep{Zhang18,Zhang19}, one can examine how each NS observable may help constrain the high-density EOS parameter space. We note that results from the direct inversion and statistical inversion using the Bayesian approach were found consistent \citep{Xie19}. As examples relevant for the present study, shown in Figure \ref{M214a} are the constant surfaces of several observables and physics conditions in the 3D $K_{\rm sym}-J_{\rm sym}-J_0$ EOS parameter space: the NS maximum mass of M=2.14 $M_{\odot}$ (green surface) or 2.01 $M_\odot$ (pink surface), the radius of canonical NS $R_{1.4}=12.83$ km (yellow surface) or $R_{1.28}=11.52$ km (orange surface) for a NS of mass 1.28 $M_{\odot}$, the dimensionless tidal deformability of canonical NS $\Lambda_{1.4}=580$ (red surface), and the causality surface (blue) on which the speed of sound equals the speed of light at the central density of the most massive NS supported by the nuclear pressure at each point with the specific EOS there \citep{Zhang19}. The red arrows point to the constant surfaces on which the specified observables have the same values while the black arrows indicate the directions satisfying the specified observational constraints. All acceptable EOSs have to support NSs at least as massive as M=2.14 $M_{\odot}$ but below the causality surface. This limits mainly the EOS space vertically, namely, the range of $J_0$ parameter, while the radii and tidal deformability provide more strict constrains horizontally, namely, the $K_{\rm sym}$ and $J_{\rm sym}$ as well as their correlations. The cross lines of these constant surfaces determines the boundaries of the 3D EOS parameter space. For example, the crossline between the causality surface and the maximum mass of M=2.14 $M_{\odot}$ sets a boundary from the lower-right side while the crossline between the causality surface and the upper limit of canonical NS radius $R_{1.4}\leq12.83$ km sets an upper-left boundary for the EOS space. It is also seen that the latter is slightly tighter than the one from using the upper limit of the tidal deformability $\Lambda_{1.4}\leq 580$ from the GW170817 data.

We now examine possible updates to the constrained 3D high-density EOS parameter space considering the results of NICER's simultaneous measurements of both the radius and mass of PSR J0030+0451. Their data were analyzed using two slightly different approaches, but giving very consistent results. In \citet{Miller2019}, the mass and radius of PSR J0030+0451 were found to be $M=1.44^{+0.15}_{-0.14}$ $M_\odot$ and $R=13.02^{+1.24}_{-1.06}$ km at 68\% confidence level, while in \citet{Riley2019}, $M=1.34^{+0.16}_{-0.15}$ $M_\odot$ and $R=12.71^{+1.19}_{-1.14}$ km as well as $M/R=0.156^{+0.010}_{-0.008}$ were obtained.  As the compactness $M/R$ is normally considered to provide tighter and additional constraints on the mass-radius relation, we use the results from \citet{Riley2019} in the following discussions while our conclusions using the results from \citet{Miller2019} are the same.
\begin{figure}
  \centering
  \includegraphics[width=8cm]{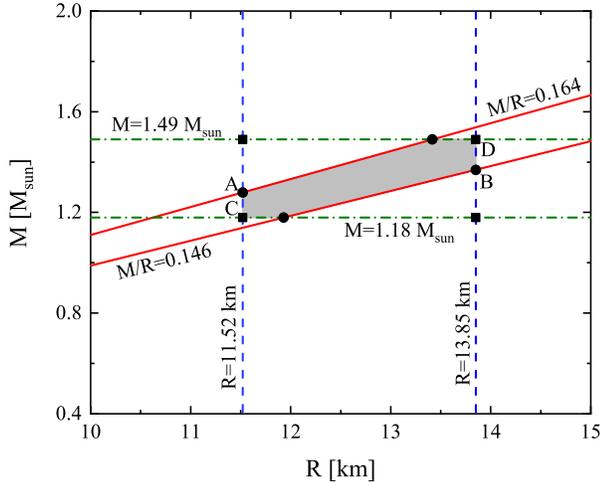}
  \caption{The observational constraints of mass $M=1.34^{+0.16}_{-0.15}$ $M_\odot$ (green dash dot lines), radius $R=12.71^{+1.19}_{-1.14}$ km (blue dash lines), and compactness $M/R=0.156^{+0.010}_{-0.008}$ (red solid lines) at 68\% confidence level from NICER observations in the mass-radius plane. The intersections between mass and radius are labeled as black squares and the ones between compactness and mass or radius are labeled as black dots. The shadowed area corresponds to the mass-radius relation meeting all constraints from the NICER observations reported in \citet{Riley2019}. }\label{NICERconstraint}
\end{figure}

\begin{figure}
\vspace{-0.5cm}
  \centering
  \includegraphics[bb=5 52 555 587, width=8cm]{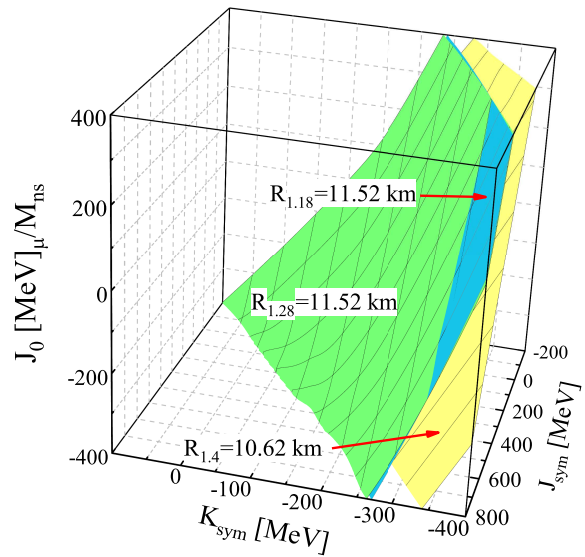}
  \caption{Lower limits on NS radius corresponding to the point A (green surface) and C (blue surface) from NICER \citep{Riley2019} shown in Figure \ref{NICERconstraint} and the lower limit (yellow surface) from analyzing quiescent low-mass X-ray binaries in \citet{Lattimer2014} in the 3D parameter space of $K_{\rm sym}-J_{\rm sym}-J_0$. The space on the left side of each surface is allowed.}\label{M214b}
\end{figure}
\begin{figure*}
  \centering
     \resizebox{0.45\textwidth}{!}{
     \includegraphics[bb=5 52 555 587]{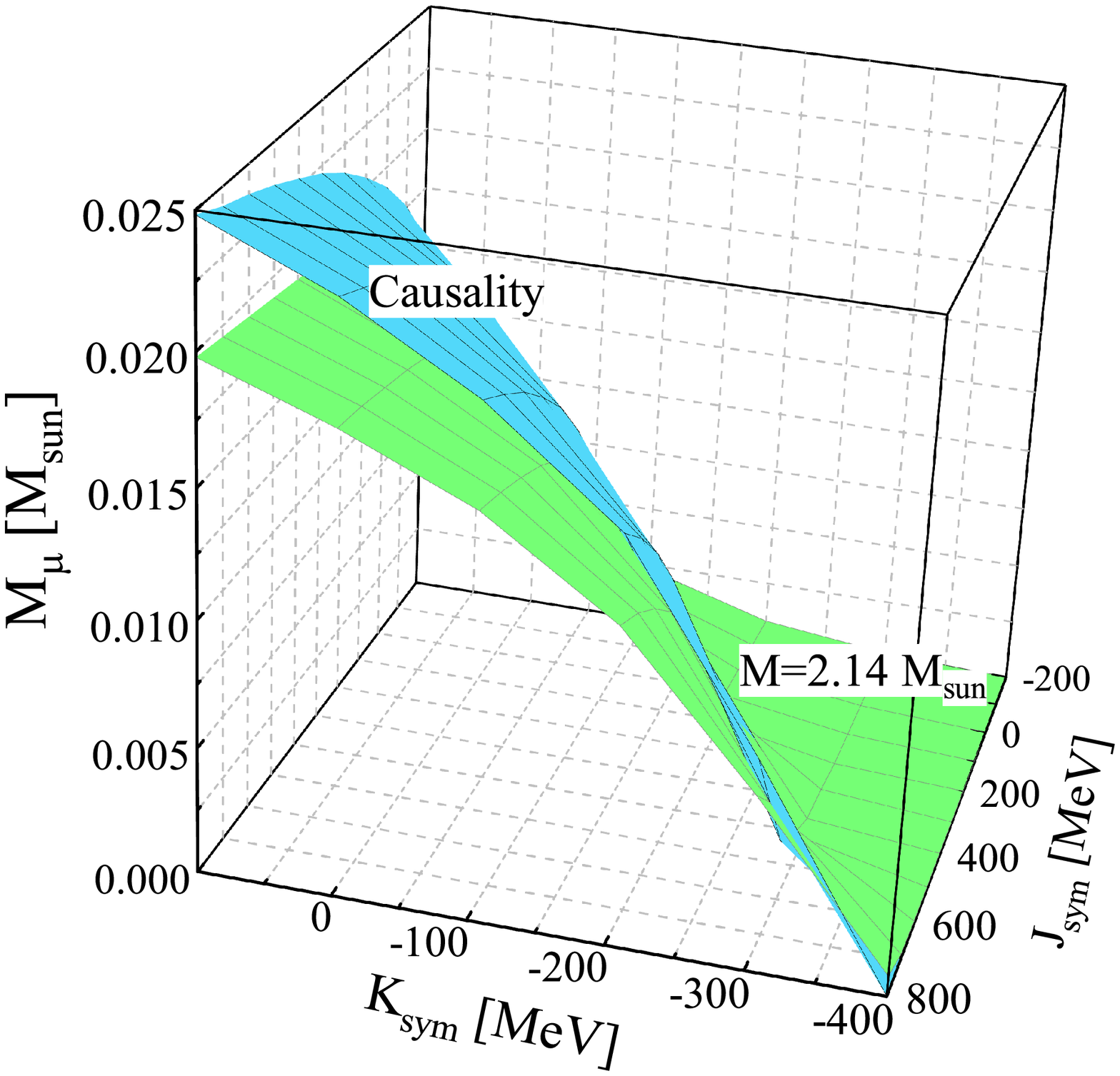}
  }
     \resizebox{0.45\textwidth}{!}{
 \includegraphics[bb=5 52 555 587]{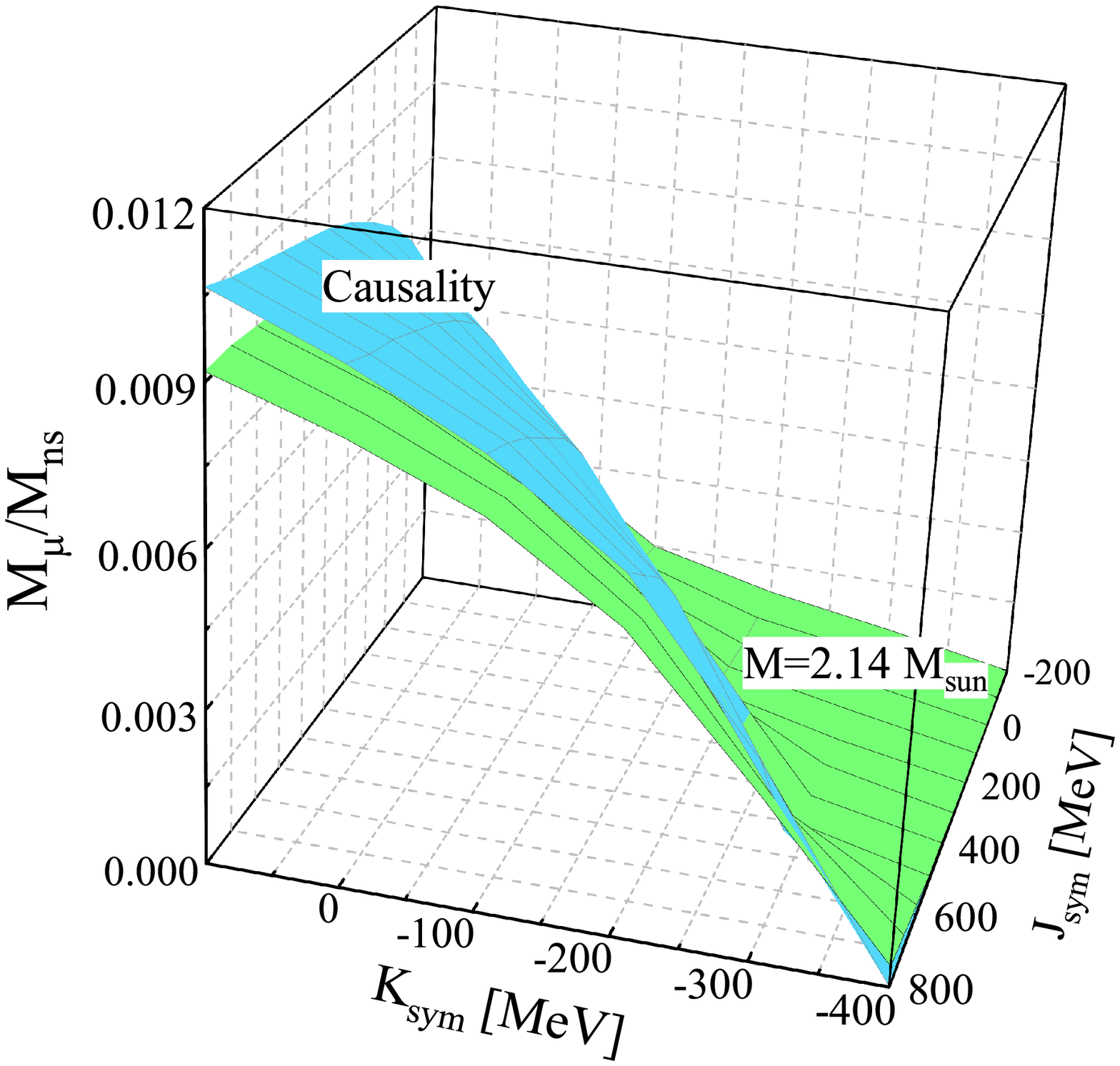}
  }
  \caption{The muon mass $M_{\mu}$ (left panel) and its mass fraction $M_{\mu}/M_{\rm NS}$ constrained by the $M=2.14 M_\odot$ (green surface) and causality condition (blue surface) in the 3D EOS parameter space of $K_{\rm sym}-J_{\rm sym}-J_0$.}\label{Mumassfraction}
\end{figure*}

The NICER results from \citet{Riley2019} are shown in the mass-radius plane in Figure \ref{NICERconstraint}. The intersections between the mass and radius are labeled as black squares and the ones between compactness and mass or radius are labeled as black dots. The shadowed area corresponds to the mass-radius relations that satisfy the constraints from NICER. As NSs with masses in the approximate range of $0.8\leq M \leq 2.0$ have almost the same radius in most calculations, the radius data provide potentially tighter and/or new constraints on the EOS parameters. Thus, the points A ($M=1.28$ $M_\odot$, $R=11.52$ km) and C ($M=1.18$ $M_\odot$, $R=11.52$ km) may provide a lower limit to the EOS parameter space, while the points B ($M=1.37$ $M_\odot$, $R=13.85$ km) and D ($M=1.49$ $M_\odot$, $R=13.85$ km) may provide an upper one. However, the latter does not provides a more tight constraint than the $R_{1.4}\leq 12.83$ km and $\Lambda_{1.4}\leq 580$ boundaries shown in Figure \ref{M214a} while they are all very close and consistent. To determine whether the point A or C provides a tighter constraint, we compare the constant surfaces of points A (green surface) and C (blue surface) from Figure \ref{NICERconstraint} as well as the lower limit  of $R_{\rm{1.4}}\geq 10.62$ km (yellow surface) from analyzing the quiescent low-mass X-ray binaries in the 3D high-density EOS parameter space in Figure \ref{M214b}. The space on the left side of each constant surface is allowed. It is clearly seen that the $R_{1.28}\geq 11.52$ km provides the tightest constraint on the EOS parameter space except at the right front corner. As the latter can be easily excluded by the causality condition, the lower limit of the radius $R_{1.28}\geq11.52$ km sets the most useful and a new constraint by building a rigid new wall on the right-back corner of the high-density EOS parameter space. Obviously, it is significantly more constraining on the EOS space than the constraint $R_{\rm{1.4}}\geq 10.62$ km from the earlier X-ray data which also suffers from some systematic errors \citep{Lattimer2014}.

In comparison with other constraints shown in Figure \ref{M214a}, however, it is seen that the $R_{1.28}\geq11.52$ km constant surface is still outside to the right of the crossline between the constant surface of NS maximum mass of M=2.14 $M_{\odot}$ and the causality surface. Consequently, the high-density EOS parameter space surrounded by the green (lower limit of NS maximum mass), yellow (upper limit of canonical NS radius), and blue surfaces (upper limit of sound speed) shown in Figure \ref{M214a} satisfy all of the latest NS observational constraints and physics requirements discussed above.  Within this allowed high-density EOS space, we shall examine next constraints on the muon contents in NSs. Before continuing, based on our examinations above we note here that simultaneous measurements of both the masses and radii of NSs more massive than the PSR J0030+0451 with NICER and/or other observatories are expected to have the great potential of dramatically further tightening the high-density EOS parameter space.

\section{Constraints on muon contents in massive neutron stars}\label{Muon-constraints}
We have discussed effects of the high-density EOS parameters on muon contents in canonical neutron stars in Section \ref{effects}. The proposed new physics associated with the muonphilic DM becomes more important and potentially easier to be detected in more muon-rich environments. It is thus interesting to study muon contents in the most massive NSs both detected so far and maximumly allowed by causality. For this purpose, we investigate here muon contents in NSs on the constant surface of M=2.14 $M_{\odot}$ and causality, respectively. As shown in Figure \ref{M214a}, these two constant surfaces are approximately perpendicular to the constant surfaces of radii and/or tidal deformability. This observation is well understood as the masses of NSs are mainly determined by the stiffness of high-density SNM EOS characterized by the $J_0$ parameter while the radii are mostly affected by the stiffness of nuclear symmetry energy characterized by the $K_{\rm sym}$ and $J_{\rm sym}$ when the $L$ parameter is fixed at its known most probable value as in our study here. We thus present here first the muon contents on the M=2.14 $M_{\odot}$ and causality surfaces, respectively, then at selected points in the allowed EOS space.
\begin{figure*}
  \centering
  \resizebox{0.9\textwidth}{!}{
  \includegraphics{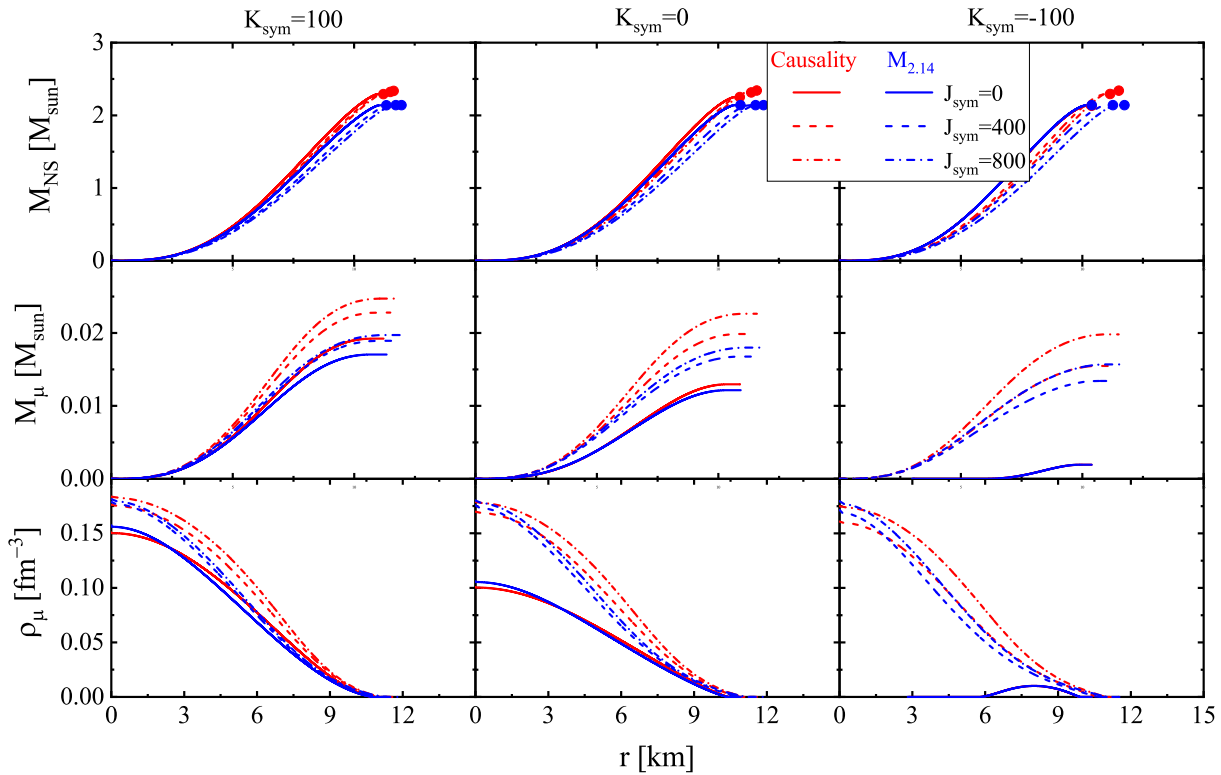}
  }
  \caption{The radial profiles of NS mass $M_{\rm NS}(r)$, muon mass $M_{\mu}(r)$, and muon density $\rho_{\mu}(r)$ with EOS parameters adopted on the surfaces of causality (red lines) and $M=2.14$ $M_\odot$ (blue lines) for $K_{\rm sym}=100$ (left panels), $0$ (middle panels), and $-100$ MeV (right panels) and $J_{\rm sym}=0$ (solid lines), $400$ (dashed lines), and $800$ (dash-dot lines) MeV, respectively. The corresponding radii $R$ for the selected parameters are labeled as red and blue dots in the upper panels.}\label{Pro}
\end{figure*}

Shown in Figure \ref{Mumassfraction} are the muon mass $M_{\mu}$ (left panel) and its mass fraction $M_{\mu}/M_{\rm NS}$ (right panel) on the M=2.14 $M_{\odot}$ and causality surfaces, respectively. By definition, the causality surface (larger $J_0$) determines the upper limit in the 3D high-density EOS space. At the front-left corner where the surface is rather flat, the muon mass $M_\mu$ and its fraction $M_\mu/M_{\rm NS}$ reach their absolute upper limits of 0.025 $M_\odot$ and 0.018, respectively. Staying on the causality surface and moving to the bottom-right corner, the combinations of $K_{\rm sym}$ and $J_{\rm sym}$ make the symmetry energy super-soft, leading to the formation of pure neutron matter without any lepton at all, thus zero muon fraction. Similarly, the muon mass and its fraction on the $M=2.14$ $M_\odot$ (smaller $J_0$) surface vary with $K_{\rm sym}$ and $J_{\rm sym}$. The muon contents on the M=2.14 $M_{\odot}$ and causality surfaces becomes closer when the NS mass 2.14 $M_{\odot}$ becomes the absolutely maximum mass allowed by causality.
\begin{figure*}
  \centering
   \resizebox{0.7\textwidth}{!}{
 \includegraphics{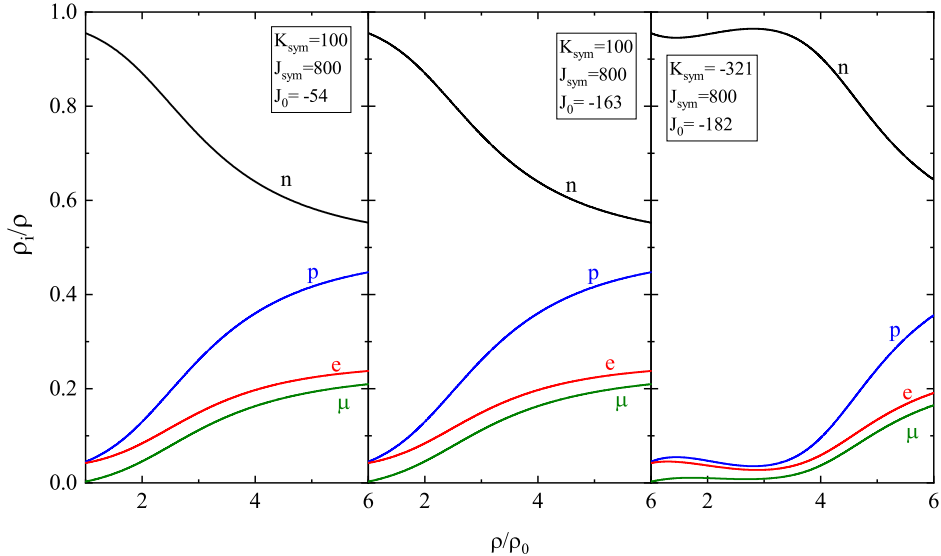}
 }
  \caption{The particle fraction $\rho_i/\rho$ as functions of reduced baryon density $\rho/\rho_0$ with EOS parameters on the causality (left), $M=2.14$ $M_\odot$ (middle) surfaces, and the cross line (right) between these two surfaces.}\label{Particle}
\end{figure*}

More detailed information about the muon contents on the M=2.14 $M_{\odot}$ and causality surfaces can be obtained. Shown in Figure \ref{Pro} are several representative radial profiles of NS mass $M_{\rm NS}(r)$, muon mass $M_\mu(r)$, and muon density $\rho_{\mu}(r)$ with selected parameters on the causality surface (red lines) and $M=2.14$ $M_\odot$ surface (blue lines) for $K_{\rm sym}=100$ (left panels), $0$ (middle panels), and $-100$ MeV (right panels) and $J_{\rm sym}=0$ (solid lines), $400$ (dashed lines), and $800$ (dash-dot lines) MeV, respectively. The corresponding radii $R$ for the selected parameters are labeled as red and blue dots in the upper panels. Compared to the $M_{\rm NS}$ and $M_\mu$ radial profiles, the muon density profile $\rho_{\mu}(r)$ is affected more apparently by the variation of nuclear symmetry energy as one expects. It is also seen that effects of varying $J_{\rm sym}$ are larger for smaller values of $K_{\rm sym}$. Overall, the muon fraction is the highest on the causality surface with the hardest symmetry energy achieved with large positive values for $K_{\rm sym}$ and/or $J_{\rm sym}$. However, when the symmetry energy is super-soft with large negative values for $K_{\rm sym}$, e.g., in the right column, the muon contents are low on both the causality and $M=2.14$ $M_\odot$ surfaces especially if the $J_{\rm sym}$ value is also relatively low, e.g., with $J_{\rm sym}=0$ (in this case the blue and red lines coincide) in the right-bottom panel. These features are consistent with what we found for canonical NSs discussed earlier.

For studying the new physics related to muonphilic DM in NSs, it is also useful to know the correlations among densities of neutrons, protons, electrons, and muons. Shown in Figure \ref{Particle} are the particle fractions $\rho_i/\rho$ (with $i=n,p,e$, and $\mu$, respectively) as functions of baryon density $\rho$ for three typical EOSs on the causality surface (left), $M=2.14$ $M_\odot$ (middle) surface, and their cross line (right), respectively. As we can see, the maximum muon fraction $\rho_{\mu}/\rho$ is around 0.2 reached at $\rho\approx 6\rho_0$ on both the causality and $M=2.14$ $M_\odot$ surfaces. In fact, it is seen clearly that the density correlations in  the left and middle windows are almost identical. This is not surprising as the two sets of EOS parameters have the same $K_{\rm sym}$ and $J_{\rm sym}$ but different $J_0$ values, namely the same symmetry energy function $E_{\rm sym}(\rho)$. The latter determines uniquely the proton (lepton) fraction in NSs at $\beta$ equilibrium as we discussed in detail in Section \ref{model}. It is the different values of $J_0$ that make one EOS stiff enough to support a NS on the causality surface while the other one having a maximum mass of $M=2.14$ $M_\odot$. When the particle fractions are examined as a function of baryon density up to $6\rho_0$ reached in both NSs, they are almost identical as one expects. We note that this study is different from examining the muon mass
$M_{\mu}$ and fraction $M_{\mu}/M_{\rm NS}$ shown in Figure \ref{Mumassfraction} as functions of $K_{\rm sym}$ and $J_{\rm sym}$ which vary the function $E_{\rm sym}(\rho)$ itself.

\section{Summary and outlook}
Motivated by the critical roles played by muons in NSs in addressing several novel physics questions associated with muonphilic DM particles proposed in the literature, we investigated the muon contents in NSs within a minimum model for NSs consisting of neutrons, protons, electrons, and muons. Considering the possible formation of negatively charged hyperons and baryon resonances that may reduce the muon fraction in NSs at $\beta$ equilibrium, our evaluations of the muon contents within the $npe\mu$ model should be considered as the most optimistic estimates useful mostly for the exploratory studies of DM particles in NSs.

Using an explicitly isospin-dependent parametric EOS for the core of NSs and the technique of solving the NS inverse-structure problems developed in our previous work, we first updated the constraints on the high-density EOS parameter space considering the latest observations of NS mass, radius, and tidal deformability including the simultaneous measurement of both the mass and radius of PSR J0030+0451 by the NICER Collaboration. We then examined the total muon mass, mass fraction and radial profile for both canonical and massive NSs within the high-density EOS space allowed by the latest observational constraints. We also studied particle fractions as a function of baryon density on both the causality surface and for the most massive NS observed so far. We found that the absolutely maximum muon mass $M_{\mu}$ and its mass fraction $M_{\mu}/M_{\rm NS}$ achieved on the causality surface are about 0.025 $M_\odot$ and 1.1\%, respectively. In the most massive NS observed so far with
M=2.14 $M_\odot$, they reduce to about 0.020 $M_\odot$ and 0.9\%, respectively. These results can be used as restricted inputs in model studies of the new physics associated with the muonphilic DM particles in NSs.

We examined in detail the individual roles of the stiffness of high-density SNM EOS characterized by the $J_0$ parameter and the high-density behavior of nuclear symmetry energy characterized by the $K_{\rm sym}$ and $J_{\rm sym}$ parameters. We found that for any NS with a given mass, the most uncertain nuclear physics input for determining the muon mass, its mass fraction, and density profile is the high-density behavior of nuclear symmetry energy. Besides new laboratory experiments with high-energy radioactive ion beams at advanced rare isotope beam facilities being built around the world, see, e.g., Refs. \citep{FRIB,LRP1,LRP2}, ongoing and planned NS observations, see, e.g., Refs. \citep{Watts19,wp1,wp2}, with advanced X-ray and/or gravitational wave detectors will all help further constrain the EOS of dense neutron-rich matter. Especially, energetic heavy-ion collisions involving highly neutron-rich nuclei in terrestrial laboratories together with simultaneous measurements of both the masses and radii of massive NSs and/or high-frequency gravitational wave signals from the merging phase of colliding NSs in space have the great potentials of finally fixing the high-density behavior of nuclear symmetry energy.

\section*{Acknowledgement} We would like to thank Wen-Jie Xie for helpful discussions. This work is supported in part by the U.S. Department of Energy, Office of Science, under Award Number DE-SC0013702, the CUSTIPEN (China-U.S. Theory Institute for Physics with Exotic Nuclei) under the US Department of Energy Grant No. DE-SC0009971, the China Postdoctoral Science Foundation under Grant No. 2019M652358, and the Fundamental Research Funds of Shandong
University under Grant No. 2019ZRJC001.


\newpage
\begin{thebibliography}{}

\bibitem[Abbott et al.(2018)]{LIGO18} Abbott, B. P., et al. 2018, \prl, 121, 161101
\bibitem[Balantekin et al.(2014)]{FRIB} Balantekin, A.B. et al. (FRIB Theory Alliance Steering Committee) 2014, Nuclear Theory at the Facility for Rare Isotope Beams (FRIB),
Mod. Phys. Lett. A 29, 1430010
\bibitem[Baillot et al.(2019)]{Baillot19} Baillot d'Etivaux, N., et al. 2019, \apj, 887, 48
\bibitem[Baran et al.(2005)]{ditoro} Baran, V., Colonna, M., Greco, V. \& Di Toro, M. 2005, Phys. Rep. \textbf{410}, 335
\bibitem[Baym et al.(1971)]{Baym71} Baym, G., Pethick, C. J., \& Sutherland, P. 1971, \apj, 170, 299
\bibitem[Bell et al.(2019)]{DM-review} Bell, N. F., Busoni, G., \&  Robles, S. 2019, \jcap, 2019, 054
\bibitem[Bilous et al.(2019)]{Bil} Bilous, A. V. et al. 2019, \apjl, 887, L23
\bibitem[Bombaci \& Lombardo(1991)]{Bom91} Bombaci, I., \& Lombardo, U. 1991, \prc, 44, 1892
\bibitem[Bogdanov et al.(2019a)]{Bog1} Bogdanov, S. et al. 2019, \apjl, 887, L25
\bibitem[Bogdanov et al.(2019b)]{Bog2} Bogdanov, S. et al. 2019, \apjl, 887, L26
\bibitem[Bogdanov et al.(2019)]{wp1} Bogdanov, S. et al, 2019, arXiv:1903.04648v1, a science white paper submitted for the Astro2020 Decadal Survey on Astronomy and Astrophysics
\bibitem[Brown \& Schwenk(2014)]{Brown14} Brown, B. A., \& Schwenk, A. 2014, \prc, 89, 011307, Erratum: [2015, \prc, 91, 049902].
\bibitem[Cai et al.(2015)]{Cai} Cai, B.J., Fattoyev, F. J., Li, B. A., \& Newton, W. G. 2015, \prc, 92, 015802
\bibitem[Cromartie et al.(2019)]{Mmax} Cromartie, H. T., et al. 2019, Nature Astronomy, 439
\bibitem[Das et al.(2020)]{Das20} Das, H.C. et al. 2020, arXiv:2002.00594
\bibitem[Danielewicz et al.(2002)]{Danielewicz02} Danielewicz, P., Lacey, R., \& Lynch, W. G. 2002, Sci., 298, 1592
\bibitem[Drago et al.(2014)]{D1}Drago, A., Lavagno, A., Pagliara, G., \& Pigato,  D. 2014, \prc, 90, 065809
\bibitem[Dror et al.(2019)]{Dror19} Dror, J. A., Laha, R., \& Opferkuch, T. 2019, arXiv:1909.12845
\bibitem[Fonseca et al.(2019)]{wp2} Fonseca, E. et al. 2019, arXiv:1903.08194v1, a science white paper submitted for the Astro2020 Decadal Survey on Astronomy and Astrophysics
\bibitem[Garani \& Heeck(2019)]{Garani19} Garani, R., \& Heeck, J. 2019, \prd, 100, 035039
\bibitem[Guillot et al.(2019)]{Gui}Guillot, S. et al. 2019, \apjl, 887, L27
\bibitem[Lattimer \& Steiner(2014)]{Lattimer2014} Lattimer, J. M., \& Steiner, A. W. 2014, Eur. Phys. J. A, 50, 40
\bibitem[Li \& Steiner(2006)]{LiSteiner}Li, B. A. \& Steiner, A.W. 2006, Phys. Lett. B, 642, 436
\bibitem[Li et al.(2008)]{LCK08} Li, B. A., Chen, L. W., \& Ko, C. M. 2008, Phys. Rep., 464, 113
\bibitem[Li \& Han(2013)]{Li13} Li, B. A., \& Han, X. 2013, Phys. Lett. B, 727, 276
\bibitem[Li et al.(2014)]{Tesym} Li, B. A., Ramos, \`{A}., Verde G., \& Vida\~{n}a, I. (Eds.) 2014, {\it Topical Issue on Nuclear Symmetry Energy}, Eur. Phys. J. A, 50, 9
\bibitem[Li \& Sedrakian(2019)a]{Armen1} Li, J. J., \& Sedrakian, A. 2019, \prc, 100, 015809
\bibitem[Li et al.(2019)]{BALI19}Li, B.A., Krastev, P.G., Wen, D.H., \& Zhang, N.B. 2019, Euro. Phys. J. A, 55, 117
\bibitem[Masuda et al.(2013)]{Masuda13} Masuda, K., Hatsuda, T., and Takatsuka, T. 2013, Prog. Theor. Exp. Phys., 2013, 073D01
\bibitem[U.S. LRP(2015)] {LRP1} {The 2015 U.S. Long Range Plan for Nuclear Science, Reaching for the Horizon,}\\
\url{https://science.energy.gov/~/media/np/nsac/pdf/2015LRP/2015_LRPNS_091815.pdf}
\bibitem[NuPECC LRP(2017)]{LRP2} {The Nuclear Physics European Collaboration Committee (NuPECC) Long Range Plan 2017, Perspectives in Nuclear Physics,}\\
\url{http://www.esf.org/fileadmin/user_upload/esf/Nupecc-LRP2017.pdf}.
\bibitem[Riley et al.(2019)]{Riley2019} Riley, T. E. et al. 2019, \apjl, 887, L21
\bibitem[Miller et al.(2019)]{Miller2019} Miller, M. C. et al. 2019, \apjl, 887, L24
\bibitem[Muller \& Serot(1995)]{Muller95} Muller, H., \& Serot, B. 1995, Phys. Rev. C, 52, 2072
\bibitem[Nakazato \& Suzuki(2019)]{Nakazato19} Nakazato, K. \& Suzuki, H. 2019, \apj, 878, 25
\bibitem[Niven (2020)]{WK} Niven, L., 2020, Wikipedia, \url{https://en.wikipedia.org/wiki/Neutron_star#Density_and_pressure}
\bibitem[Negele \& Vautherin(1973)]{Negele73} Negele, J. W., \& Vautherin, D. 1973, \nphysa, 207, 298
\bibitem[Oertel et al.(2017)]{Oertel17} Oertel, M., Hempel, M., Kl\"{a}hn, T., \& Typel, S. 2017, Rev. Mod. Phys., 89, 015007
\bibitem[Oppenheimer \& Volkoff(1939)]{Oppenheimer39} Oppenheimer, J., \& Volkoff, G. 1939, Phys. Rev., 55, 374
\bibitem[Piekarewicz(2010)]{Piekarewicz10} Piekarewicz, J. 2010, JPhG, 37, 064038
\bibitem[Provid\^{e}ncia et al.(2019)]{Pro19} Provid\^{e}ncia et al., C. 2019, Front. Astron. Space Sci., 26 March\\
\url{ https://doi.org/10.3389/fspas.2019.00013}
\bibitem[Raaijmakers et al.(2019)]{Raa} Raaijmakers, G. et al. 2019, \apjl, 887, L22
\bibitem[Ribes et al.(2019)]{Ramos}Ribes, P., Ramos, \`{A}., Tolos, L., Gonzalez-Boquera, C., \& Centelles, M. 2019, \apj, 883, 168
\bibitem[Sahoo et al.(2018)]{Sahoo}Sahoo, H. S., Mitra, G., Mishra, R., Panda, P.K., \& Li, B.A. 2018, \prc, 98,  045801
\bibitem[Shlomo et al.(2006)]{Shlomo06} Shlomo, S., Kolomietz, V. M., \& Col\`{o} G. 2006, Eur. Phys. J. A, 30, 23
\bibitem[Steiner et al.(2010)]{Steiner10} Steiner, A. W., Lattimer, J. M., \& Brown, E. F. 2010, \apj, 722, 33
\bibitem[Takeda et al.(2018)]{Kim} Takeda, Y., Kim, Y., \& Harada,  M. 2018, \prc, 97, 065202
\bibitem[Tews et al.(2017)]{Tews17} Tews, I., Lattimer, J. M., Ohnishi, A., \& Kolomeitsev, E. E. 2017, \apj, 848, 105
\bibitem[Vida\~{n}a et al.(2003)]{Isaac1} Vida\~{n}a, I, Bombaci, I., Polls, A., \& Ramos, \`{A}. 2003, Astronomy \& Astrophysics, 399, 687
\bibitem[Vida\~{n}a(2016)]{Isaac2} Vida\~{n}a, I. 2016, Journal of Physics: Conference Series, 668, 012031
\bibitem[Watts (2019)]{Watts19} Watts, A.L. 2019, in AIP Conf. Proc. 2127, Xiamen-CUSTIPEN Workshop on the Equation of State of Dense Neutron-Rich Matter in the Era of
Gravitational Wave Astronomy, ed. A. Li, B.-A. Li, \& F. Xu (Melville, NY: AIP), 020008
\bibitem[Weber (2005)]{Weber05} Weber, F. 2005, Progress. in Part. and Nucl. Phys., 54, 193
\bibitem[Xie \& Li(2019)]{Xie19}Xie, W. J., \& Li, B.A. 2019, \apj, 883, 174
\bibitem[Xu et al.(2000)]{Xu00} Xu, H.S. et al. 2000, Phys. Rev. Lett. 85, 716
\bibitem[Zhang et al.(2017)]{Zhang17} Zhang, N. B., Cai, B. J., Li, B. A., Newton, W. G., \& Xu, J. 2017, Nucl. Sci. Tech., 28, 181
\bibitem[Zhang et al.(2018)]{Zhang18} Zhang, N. B., Li, B. A., \& Xu, J. 2018, \apj, 859, 90
\bibitem[Zhang \& Li(2019a)]{Zhang19} Zhang, N. B., \& Li, B. A. 2019, Eur. Phys. J. A, 55, 39
\bibitem[Zhang \& Li(2019b)]{Zhang19a} Zhang, N. B., \& Li, B. A. 2019, JPhG, 46, 014002
\bibitem[Zhang \& Li(2019c)]{Zhang19b}Zhang, N. B., \& Li, B. A. 2019, \apj, 879, 99
\bibitem[Zhou et al.(2019)]{LWChen19} Zhou, Y., Chen, L. W., \& Zhang, Z. 2019, \prd, 99, 121301
\bibitem[Zhou \& Chen(2019)]{YZhou19} Zhou, Y., \& Chen, L. W. 2019, \apj, 886, 52
\bibitem[Zhu et al.(2016)]{AngLi} Zhu, Z. Y., Li,  A., Hu, J. N., \& Sagawa, H. 2016, \prc, 94, 045803
\end{thebibliography}
\end{document}